\documentclass{ws-p8-50x6-00}

\newcommand{\bcen}{\begin{center}}
\newcommand{\ecen}{\end{center}}

\newcommand{\beq}{\begin{equation}}
\newcommand{\eeq}{\end{equation}}
\newcommand{\beqn}{\begin{eqnarray}}
\newcommand{\eeqn}{\end{eqnarray}}

\newcommand{\PRC}[3]{{\it Phys. Rev.} C {\bf #1}, #2 (#3)}
\newcommand{\EPJ}[3]{{\it Eur. Phys. J.} C {\bf #1}, #2 (#3)}
\newcommand{\PR}[3]{{\it Phys. Rev.} {\bf #1}, #2 (#3)}
\newcommand{\NPA}[3]{{\it Nucl. Phys.} A {\bf #1}, #2 (#3)}
\newcommand{\PPNP}[3]{{\it Prog. Part. Nucl. Phys.} {\bf #1}, #2 (#3)}

\begin{document}

\title{Vector Meson Decay of Baryon Resonances}
\author{U. Mosel and M. Post$^{\star}$}
\address{Institut f\"ur Theoretische Physik, Universit\"at 
Giessen, D-35392 Giessen \\ 
$^{\star}$Email: Marcus.Post@theo.physik.uni-giessen.de}


\maketitle

\abstracts{We investigate the coupling of vector mesons with nucleons to
nucleon resonances in an isospin-selective VMD approach and explore the
in-medium properties of vector mesons. }


\section{Introduction} One of the evidences for the discovery of a new
state of matter that was quoted by the CERN press release in 2000 was the
result of the CERES experiment\cite{ceres1} that showed an excess of
dileptons at invariant masses below the vector meson mass. In order to
understand this phenomenon it is essential to consider the effect of
conventional hadronic interactions on the vector mesons. We emphasize the
relevance of the excitation of baryon resonances in this context. To this
end we analyze the decay of nucleon resonances into vector mesons within
an isospin selective VMD model. In the isovector channel we compare the
results to fits to the hadronic $N\rho$ decay width. In the isoscalar
channel we then predict the $\omega$ coupling strength.


\section{VMD Analysis of the Electromagnetic Resonance Decay}

Vector Meson Dominance (VMD)\cite{williams}, a theory which describes
photon-hadron interactions exclusively
in terms of vector meson-hadron interactions,
relates the hadronic coupling strength of resonances to vector mesons 
$f_{RN\rho(\omega)}$ and the isoscalar and isovector part of the 
photon-coupling:
\beq
\label{vmdcoupl}
f_{RN\omega} = g_s \, m_\omega \, \frac{2\,g_\omega}{e} \quad,\quad 
f_{RN\rho} = g_v \, m_\rho \, \frac{2\,g_\rho}{e}  \quad. 
\eeq
As values for $g_\rho$ and $g_\omega$ -- the coupling strengths 
of $\rho$ and $\omega$
meson to the photon -- we take $g_\rho = 2.5$ and  $g_\omega = 8.7$ (ref. 
\cite{williams}). 
The isoscalar and isovector  
coupling strength of the resonance to the $N\gamma$ system 
is given by $g_s$ and $g_v$, respectively, see Eq. \ref{isospin}.
Thus VMD gives access to both $f_{RN\omega}$ and $f_{RN\rho}$,
if it is possible to obtain $g_s$ and $g_v$ 
from experimental data.
In order to achieve this goal, the coupling has to be 
decomposed into an isoscalar and an isovector part,
which is readily done by constructing suitable linear combinations
of proton- and neutron-amplitudes\cite{pm}.

The isospin part $I$ of the electromagnetic coupling is given by:
\beqn
\label{isospin}
I = \chi_R^{I\,\dagger}\, \left( g_s + g_v\, \tau_3 
 \right)\, \chi^I_N  
\eeqn
For simplicity we restrict ourselves here to the case of isospin 
1/2; the case of isospin 3/2 contains additional Clebsch-Gordan 
coefficients which are given in ref.\cite{pm}.
The spinors $\chi_R^I$ and $\chi_N^I$ represent resonance and
nucleon iso spinors and $I_R$ denotes the isospin of the resonance.
$\tau_3$ refers to the Pauli matrix.  
From the structure of the isospin coupling $I$ it follows that 
the linear combinations
\beq
\label{scavec}
{\cal M}_{s/v} = \frac{1}{2}\, \left( {\cal M}_p \pm {\cal M}_n \right) 
\eeq
are proportional to $g_s$ and $g_v$ respectively.

At the pole-mass of the resonance the helicity amplitudes 
$A_{\frac12}^{p/n}$ and $A_{\frac32}^{p/n}$ are known from experiment.
Therefore also $g_s$ and $g_v$ are determined except for a normalization 
factor. We calculate this factor by introducing
the $\gamma$-width  $\Gamma_{s/v}^{\gamma}$,
defined in terms of the helicity amplitudes
$A_{s/v}$ as follows\cite{pdg}:
\beq
\label{gammawidth}
\Gamma_{s/v}^{\gamma}(m_R) = \frac{{\bf q_{cm}}^2}{\pi}\,
\frac{2 m_N}{(2j_R+1)m_R}\,\left(|A_{\frac12}^{s/v}|^2 + 
|A_{\frac32}^{s/v}|^2 \right)\quad, 
\eeq
with $j_R$ and $m_R$ denoting spin and pole-mass of the resonance
and ${\bf q_{cm}}$ the cm-momentum of the photon.
Clearly, $\Gamma_{s/v}^{\gamma}$ can also be expressed 
using Feynman amplitudes:
\beq
\label{gammawidth2}
\Gamma_{s/v}^{\gamma}(k^2) 
= \frac{1}{(2j_R+1)}\frac{{\bf q_{cm}}}{8\,\pi\,k^2}\, 
|{\cal M}_{s/v}|^2 \quad, \eeq
where $\sqrt{k^2}$ is the invariant mass of the resonance.
After summing over the photon polarizations,
$|{\cal M}_{s/v}|^2$ assumes the following form:
\beq
|{\cal M}_{s/v}|^2 = 4\,m_N\,m_R\,\kappa \,g_{s/v}^2 \,q^2 \, F(k^2) \quad.
\eeq
The formfactor $F(k^2)$ at the $RN\gamma$ vertex is taken from 
ref.\cite{feupho}.
The numerical factor $\kappa$ depends on the
quantum numbers of the resonance (ref.\cite{pm}).
The two expressions Eqs. \ref{gammawidth} and \ref{gammawidth2}
can now be equated allowing to solve for $g_{s/v}$:
\beq
\label{resg}
g_{s/v}^2 = \frac{4}{\kappa} \, \frac{|A_{\frac12}^{s/v}|^2 +
|A_{\frac32}^{s/v}|^2}{{\bf q_{cm}}} 
\eeq
In this way it is possible to obtain $g_{s/v}$ from helicity amplitudes. 
The hadronic couplings $f_{RN\omega(\rho)}$ 
are then readily deduced from the VMD relation Eq. \ref{vmdcoupl}. 
The corresponding values are listed in Tables \ref{vmdrho} and \ref{vmdom}.



\section{The $\rho$ Meson}

\subsection{VMD in the Isovector Channel}

In this section we investigate the applicability of VMD for the
isovector channel of the resonance decay.
For the helicity amplitudes we use different parameter sets in order to
provide an estimate for the experimental uncertainties entering this analysis.
They are taken from Arndt {\it et al}\cite{arndt} and Feuster 
{\it et al}\cite{feupho}. The $\rho$ decay widths are taken from 
the analysis of Manley {\it et al}\cite{man2}.

In Table \ref{vmdrho} the results for the coupling constants and the
corresponding error-bars are given. As a general tendency, VMD works well
within a factor of two.  This can be seen particularly well in the case of
the $D_{13}(1520)$ and the $F_{15}(1680)$, which are the most prominent
resonances in photon-nucleon reactions, and whose $\rho$ decay widths are
also well under control.  Note in particular the large coupling constant
for the $D_{13}(1520)$ in the hadronic fit which is possible only because
of the large width of the $\rho$ meson\cite{pp98,plm}.

\begin{table}[b]
\bcen
\begin{tabular}{|l||c|c|c|} \hline
& $f_{RN\rho}$(Arndt) & $f_{RN\rho}$(Feuster) & $f_{RN\rho}$(Manley) \\ \hline
$D_{13}(1520)$ & 3.44$\pm$0.18 &  2.67 & 6.67$\pm$0.78 \\ \hline
$S_{31}(1620)$ & 0.89$\pm$0.42 &  0.10 & 2.14$\pm$0.30 \\ \hline
$S_{11}(1650)$ & 0.70$\pm$0.08 &  0.59 & 0.47$\pm$0.19 \\ \hline
$F_{15}(1680)$ & 3.48$\pm$0.39 &  ---  & 6.87$\pm$1.57 \\ \hline
$D_{33}(1700)$ & 3.96$\pm$0.77 &  3.68  & 1.962$\pm$0.67 \\ \hline
$P_{13}(1720)$ & 0.25$\pm$0.42 &  0.93 & 13.17$\pm$3.35 \\ \hline 
$F_{35}(1905)$ & 2.47$\pm$0.55 &  --- & 17.97$\pm$1.14 \\ \hline
$P_{33}(1232)$ & 13.40$\pm$0.2 & 11.96 & --- \\ \hline
\end{tabular}
\caption{\label{vmdrho} Results for $f_{RN\rho}$ from a
VMD analysis (1st and 2nd column) and a hadronic fit (3rd column).}
\ecen
\end{table}

For the $P_{13}(1720)$ and the $F_{35}(1905)$ resonances
VMD is off by an order 
of magnitude. We argue that this mismatch does not necessarily 
indicate a failure of VMD, but can be traced back to the unsatisfactory
experimental information on these two resonances. Neither the 
helicity amplitudes nor the partial $N\rho$ width are well determined
from experiment\cite{feupho,arndt,man2}.
Obviously, the extraction of the resonance parameters
is very complicated and might be sensitive  
to the details of the underlying theoretical model, such as the treatment
of the non-resonant background. 
For a conclusive VMD analysis of these
resonances it is therefore mandatory to enlarge the data base and to 
describe hadron- and photoinduced reactions within one and the same 
analysis.

We conclude that VMD works remarkably well in the isovector channel.
Therefore, our approach should yield reasonable predictions 
of the unknown coupling constants $f_{RN\omega}$. 

\subsection{The $\rho$ Spectral Function in Nuclear Matter}

Using the coupling constants obtained from the hadronic $N\rho$ 
decay width of the resonance, we calculate the spectral function 
$A_\rho^{T/L}(\omega,{\bf q})$ of 
the $\rho$ meson in nuclear matter\cite{pp98,plm} at density $\rho = 
\rho_0$. 
It is defined as:
\beq
A_\rho^{T/L}(\omega,{\bf q}) = \frac{1}{\pi}\frac{\mbox{Im } 
\Sigma^{T/L}(\omega,{\bf q})}
{(\omega^2-{\bf q}^2 - m_\rho^2+\mbox{Re } \Sigma^{T/L}(\omega,{\bf q}))^2+
\mbox{Im } \Sigma^{T/L}(\omega,{\bf q})^2}   \qquad . 
\eeq
Note that in nuclear matter transverse and longitudinal modes -- denoted 
by T and L, respectively --  have to be treated independently. 
Here $\omega$ and ${\bf q}$ denote energy and momentum relative to the
rest frame of nuclear matter.
The selfenergy $\Sigma^{T/L}({\omega,{\bf q}})$ is a sum of vacuum
and in-medium contributions. The vacuum part is given by the 
2-pion decay mode and we estimate the in-medium part within the 
low-density approximation:
\beq
\label{lowdens}
\Sigma_{med}^{T/L}(\omega,{\bf q}) = 
\frac{1}{8 m_N}\,\rho_N\,T_{tot}^{T/L}(\omega,{\bf q}) \quad .
\eeq 

\begin{figure}[h]
\bcen
\parbox{7cm}{
\epsfig{file=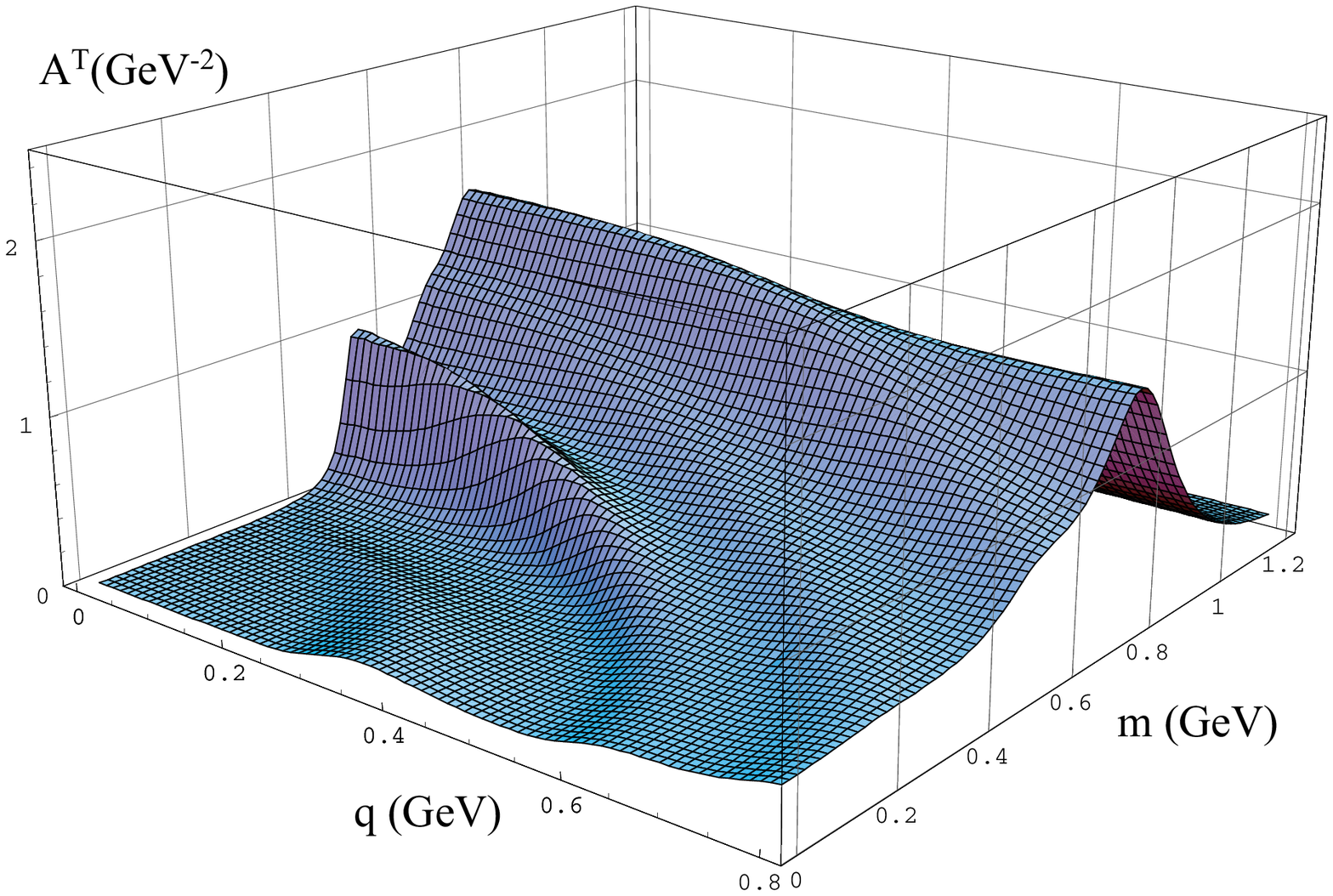,width=7cm}}\\ \parbox{7cm}{
\epsfig{file=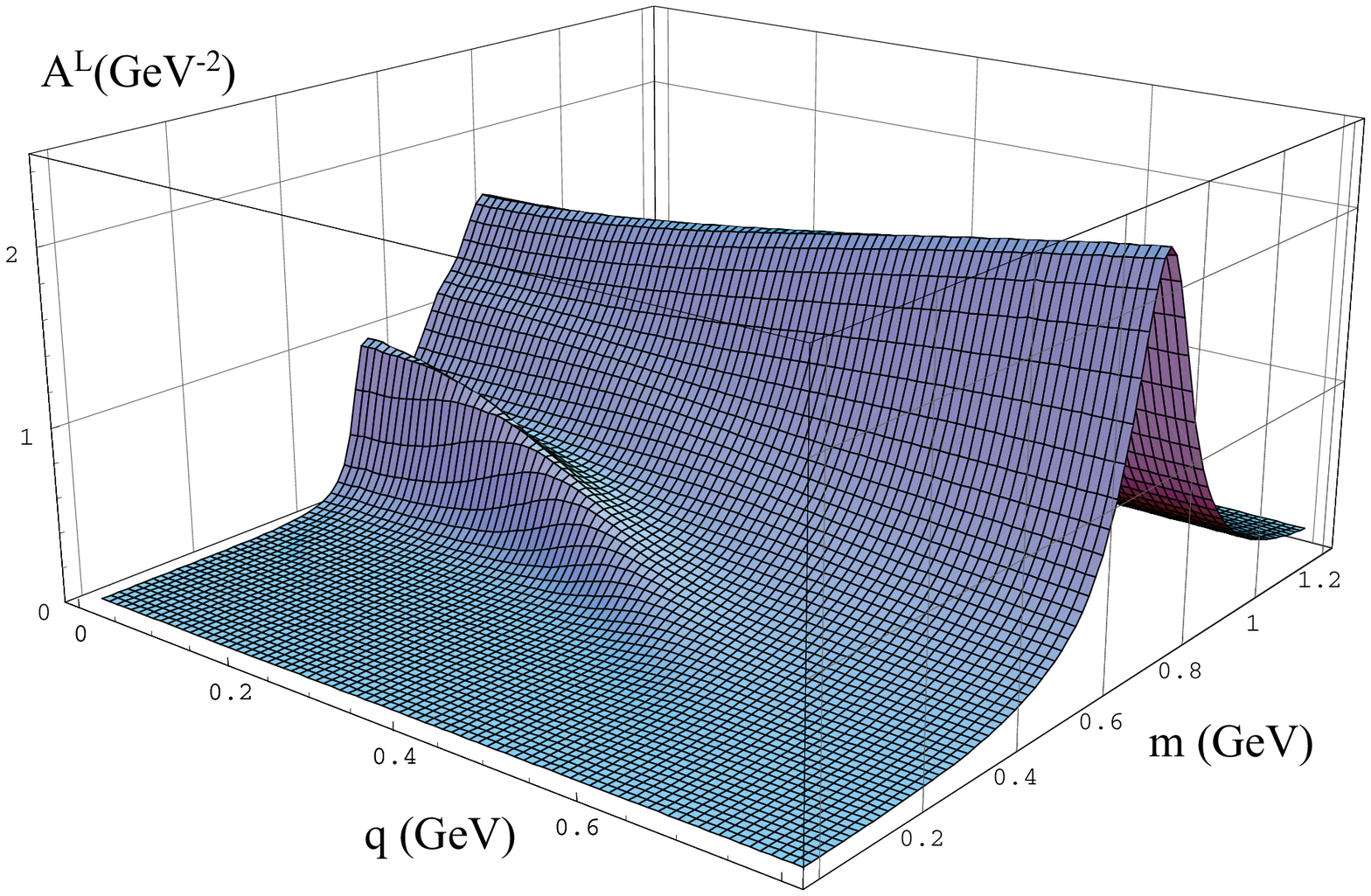,width=7cm}}
\caption{
\label{specrho}Top: $A_\rho^T$ as a function of invariant mass
$m$ and 3-momentum ${\bf q}$. Bottom: Same for $A_\rho^L$.}
\ecen
\end{figure}

The main quantity entering this expression is the $\rho\,N$ forward
scattering amplitude $T_{tot}^{T/L}$.
In Fig. \ref{specrho}
we show the results for $A_\rho^{T/L}(\omega,{\bf q})$. They 
highlight an important consequence of the strong coupling of the 
$D_{13}(1520)$ to the $N\rho$ system, namely the strong modification
of the $\rho$ spectral function in nuclear matter. In particular at
low momenta the mass spectrum is dominated by the excitation of a
$D_{13}(1520$), leading to a substantial shift of spectral strength
down to lower invariant masses. This, of course, is of relevance for the
interpretation of the CERES data. At large momenta the $P_{13}(1720)$ and the 
$F_{35}(1905)$ dominate the spectrum.
The predominant feature is the
different modification of transverse and longitudinal $\rho$ mesons.



\section{The $\omega$ Meson}

\subsection{VMD in the Isoscalar Channel}
In this section we present our results for the coupling constants
$f_{RN\omega}$ and discuss 
their compatibility with experimental information
obtained from pion- and photon-induced $\omega$-production cross sections.

All nucleon resonances with $j_R < \frac72$, for which helicity amplitudes
have been extracted, are included. Thus we consider only one resonance above
the $N\omega$ threshold in our analysis, namely the $D_{13}(2080)$.
We use again the helicity amplitudes from Arndt {\it et al}\cite{arndt} 
and Feuster {\it et al}\cite{feupho}and consider also the PDG 
estimates\cite{pdg}.
The corresponding results for $f_{RN\omega}$ 
together with the error-bars are given in 
Table \ref{vmdom}. We find a strong coupling to the
$N\,\omega$ channel in the $S_{11}$, $D_{13}$ and $F_{15}$
partial waves; especially the $S_{11}(1650)$, the $D_{13}(1520)$ and the
$F_{15}(1680)$ resonances show a sizeable coupling strength to this
channel.

\begin{table}[h]
\bcen
\begin{tabular}{|l||c|c|c|} \hline
& $f_{RN\omega}$(Arndt) & $f_{RN\omega}$(Feuster) & $f_{RN\omega}$(PDG) 
\\ \hline
$S_{11}(1535)$ & 1.27$\pm$1.58  & 1.36 & 0.76$\pm$1.23  \\ \hline
$S_{11}(1650)$ & 1.59$\pm$0.29  & 0.56 & 1.12$\pm$1.09  \\ \hline
$D_{13}(1520)$ & 2.87$\pm$0.76  & 2.28 & 3.42$\pm$0.87  \\ \hline
$D_{13}(1700)$ & $---$        & 1.88 & 0.65$\pm$2.76  \\ \hline
$D_{13}(2080)$ & $---$        &  $---$ & 1.13$\pm$1.46  \\ \hline
$P_{11}(1440)$ & 0.61$\pm$0.68  & 1.26 & 0.85$\pm$0.48   \\ \hline
$P_{11}(1710)$ & 0.14$\pm$0.85  & 0 & 0.20$\pm$1.02  \\ \hline
$P_{13}(1720)$ & 0.29$\pm$1.30  & 2.18 & 1.79$\pm$3.18  \\ \hline
$F_{15}(1680)$ & 6.89$\pm$1.38  & $----$ & 6.52$\pm$1.49  \\ \hline 
\end{tabular}
\ecen
\caption{\label{vmdom}VMD predictions for the coupling strength
$f_{RN\omega}$.}
\end{table}

It is noteworthy that the resonances with the largest coupling 
are well below the $N\omega$ threshold.
Subthreshold resonances in the $N\omega$ channel 
are also reported elsewhere\cite{frilu,brri}. 
The coupled-channel analysis\cite{frilu}
of $\pi\,N$ scattering enforces 
resonant structures in the $N\omega$ channel, in particular 
in the $S_{11}$ and $D_{13}$ partial waves.
However, the coupling strength extracted in their analysis is 
$f_{RN\omega} \approx 6.5$, nearly twice as large as our value.
In the quark model calculation of ref.\cite{brri} a value of about 
$f_{RN\omega} \approx 2.6$ is found,
which is surprisingly close to our result. 

The quality of the VMD predictions can further be tested 
by a comparison with experimental data on the reactions
$\pi^-\,p \rightarrow \omega\,n$ and $\gamma\,p \rightarrow \omega\,p$.
Comparison with data allows also to 
discuss the results for the $D_{13}(2080)$, the only resonance in our
analysis above threshold. 
We find for  this resonance
an $\omega$ decay width of about $70$ MeV  and argue 
that its contribution to both reactions is too small to be seen in experiment. 
As a first approximation, we take the full production amplitude
as an incoherent sum of {\it Breit-Wigner} type amplitudes,
describing $s$-channel contributions.

\begin{figure}[t] 
\bcen
\epsfig{file=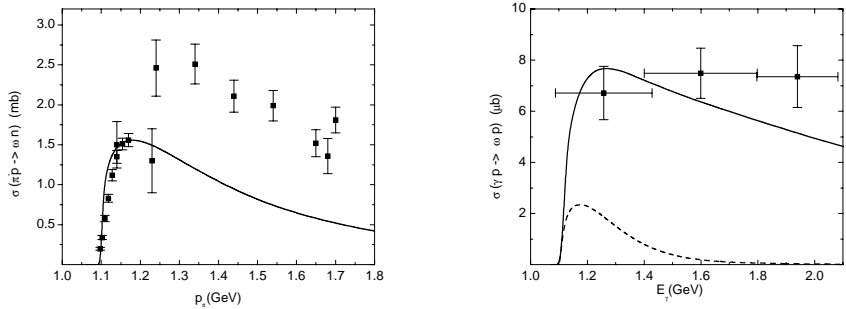,width=12cm}
\caption{
\label{pi_gamtot} Left: Total cross-section for the reaction $\pi^-\,p 
\rightarrow \omega\,n$. Right: Total cross-section for the reaction 
$\gamma\,p \rightarrow \omega\,p$ with resonance contribution only
(dashed line) and added pion-exchange (solid line).}
\ecen
\end{figure}

The results are shown in Fig. \ref{pi_gamtot}.
The data for the $\pi$-induced reaction are taken 
from ref.\cite{landolt} and we use the photoproduction data of 
ref.\cite{abbhhm}.
The cross-section for $\pi^-p \rightarrow \omega n$ is 
reproduced rather well near threshold. 
This is in agreement with the findings of ref.\cite{frilu}, where
a satisfactory description of this process around
threshold in terms of near or subthreshold resonances is presented. 
This suggests that the excitation of subthreshold resonances
constitutes an essential ingredient to the production mechanism. 
The contribution coming from the only resonance above threshold -- the 
$D_{13}(2080)$ -- is about $0.1$ mb, roughly
$10\%$ of the total cross-section. 

On the other hand, the photoproduction data cannot be saturated within
the resonance model; this gives little hope to find
the $D_{13}(2080)$ in this reaction.
However, adding the contribution from $\pi^0$-exchange yields a qualitative
explanation of the data over the energy range under consideration. 

Overall it seems that the predictions of the resonance model
are in reasonable agreement with the data and can be viewed as a confirmation
of the VMD analysis.

\subsection{The $\omega$ Spectral Function in Nuclear Matter}

\begin{figure}[t] 
\bcen
\epsfig{file=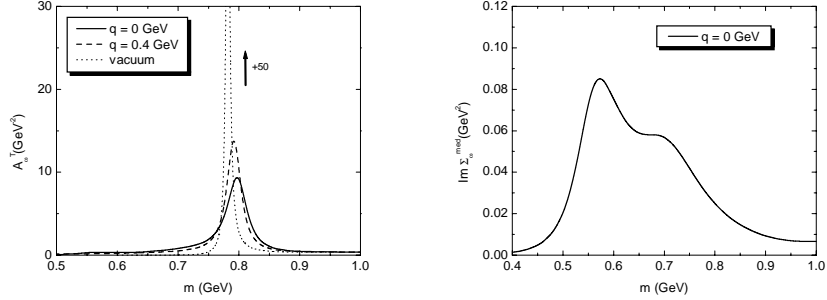,width=12cm}
\caption{
\label{specself} Left: $A_\omega^{T/L}$ at momenta $0$ GeV (straight) and
$0.4$ GeV (dashed). For comparison 
also the vacuum result is shown (dotted). Right: The imaginary part
of the $\omega$ selfenergy.
}
\ecen
\end{figure}

Within the same formalism as for the $\rho$ meson we investigate the
effects of resonance-excitation on the properties of $\omega$ mesons
in nuclear matter. 
Again the calculations are performed at $\rho=\rho_0$.
We find a broadening of the $\omega$ meson of about $50$ MeV and a
repulsive mass shift of roughly $20$ MeV, see Fig. \ref{specself}.
These findings are in surprising agreement with those of various
other groups\cite{frilu,klingl,effe}. The $\omega$ meson
thus is much less modified in nuclear matter than
the $\rho$ meson, which follows within our model from the 
much smaller coupling constants. As can be seen in Fig. \ref{specself}
the in-medium effects are most pronounced at small momenta.

\section{Conclusions}

We have demonstrated that through the excitation 
of baryon resonances the in-medium
spectral functions of vector mesons receive a substantial shift
of spectral strength down to lower invariant masses. These 
effects play a key role in understanding the in-medium properties 
of vector mesons.  

\section{Acknowledgments}
This work was supported by DFG and BMBF.

\end{document}